\documentclass[aps,prb,reprint,superscriptaddress,a4paper,twocolumn]{revtex4}
\usepackage[utf8]{inputenc}
\usepackage[T1]{fontenc}
\usepackage{amsmath}
\usepackage{graphicx}
\usepackage{dcolumn}
\usepackage{bm}
\usepackage{epsfig}
\usepackage{
float} 
\usepackage{color}
\usepackage{tikz}
\usepackage{makecell}
\usepackage{comment}
\usepackage{amssymb}
\usepackage{color} 
\usepackage{soul}
\usepackage[normalem]{ulem}
\usepackage{hyperref}
\usepackage{setspace}  
\makeatletter
\def\maketitle{
\@author@finish
\title@column\titleblock@produce
\suppressfloats[t]}
\makeatother 
\newcommand{\CYU}{CY Cergy Paris Université, CEA, LIDYL, 91191 Gif-sur-Yvette, France}
\newcommand{\CEA}{Université Paris-Saclay, CEA, LIDYL, 91191, Gif-sur-Yvette, France}
\newcommand{\Warwick}{Department of Physics, University of Warwick, Coventry, CV4 7AL, UK}
\newcommand{\NTC}{University of West Bohemia, New Technologies Research Centre, 301 00 Plzen, Czech Republic}
\newcommand{\MaxIV}{MAXIV Laboratory, Lund University, P.O. Box 118, 22100 Lund, Sweden}
\newcommand{\Solaris}{National Synchrotron Radiation Centre SOLARIS Jagiellonian University Czerwone Maki 98, Krakow 30-392, Poland}

\begin{document}

\title{  
ARPES studies of Hf(0001) surface: flat bands formation in the dice lattice}

\author{
Laxman Nagireddy}
\email{These authors contributed equally}
\affiliation{\CYU} 
\affiliation{\CEA}
\affiliation{\Warwick}

\author{Saleem Ayaz Khan}
\email{These authors contributed equally}
\affiliation{\NTC}

\author{Maria Christine Richter}
\affiliation{\CYU} 
\affiliation{\CEA}

\author{Olivier Heckmann}
\affiliation{\CYU} 
\affiliation{\CEA}

\author{Mauro Fanciulli}
\affiliation{\CYU
} 
\affiliation{\CEA}
\affiliation{\NTC}

\author{Natalia Olszowska}
\affiliation{\Solaris}

\author{Marcin Rosmus}
\affiliation{\Solaris}

\author{Weimin Wang}
\affiliation{\MaxIV}

\author{Laurent Nicola\"i}
\affiliation{\NTC} 

\author{Ján Minár}
\email{jminar@ntc.zcu.cz} 
\affiliation{\NTC}

\author{Karol  Hricovini}
\email{karol.hricovini@cyu.fr} 
\affiliation{\CYU} 
\affiliation{\CEA}
\affiliation{\NTC}

\date{\today}
\newcommand{\Karol}[1]{\textcolor{red}{\textbf{#1}}}
\newcommand{\Laxman}[1]{\textcolor{red}{\textbf{#1}}}
\newcommand{\Saleem}[1]{\textcolor{blue}{\textbf{#1}}}
\newcommand{\Mauro}[1]{\textcolor{violet}{\textbf{#1}}}
\newcommand{\Laurent}[1]{\textcolor{orange}{\textbf{#1}}}
\newcommand{\Christine}[1]{\textcolor{green}{\textbf{#1}}}
\newcommand\christineout{\bgroup\markoverwith{\textcolor{violet}{\rule[0.5ex]{2pt}{0.4pt}}}\ULon}
\newcommand\mauroout{\bgroup\markoverwith{\textcolor{violet}{\rule[0.5ex]{2pt}{0.4pt}}}\ULon}

\begin{abstract}

 We present the first electronic structure measurements of the Hf(0001) single-crystal surface using angle-resolved photoemission spectroscopy (ARPES). The ARPES results are supported by theoretical calculations performed using the full-potential linearized augmented plane wave (FLAPW) method and the Korringa-Kohn-Rostoker (KKR) Green function method.  In addition to insight into the electronic structure of Hf(0001), our results reveal the impact of surface contamination, particularly oxygen and carbon, on the predicted surface state. Moreover, we observe a flat band induced by both, the presence of oxygen and the dice structure of the surface. The orbital texture of Hf bands is confirmed by linear dichroism studies.

\end{abstract}

\maketitle

\section{Introduction}

Hafnium, group IV 5\textit{d} transition metal, was predicted by Mendeleev but identified only in 1923 and is the second-last stable element to be discovered just before rhenium, in 1925. 
In industrial applications Hf alloys are used, together with tungsten, in filaments and electrodes, in control rods in nuclear power plants, due to its large neutron capture cross section or in special alloys as, for example, in the main engine of the Apollo Lunar Modules. 
In the semiconductor industry, from 2007, HfO$_2$ began to replace SiO$_2$ and
SiON \cite{robertson2015high} as insulating barrier in field-effect transistors (FETs)  due to its high electrical permittivity and the discovery of ferroelectric properties in doped or alloyed HfO$_2$ \cite{boscke2011ferroelectricity} sparked a significant interest in its employment in ferroelectric FETs. Although extensive research has been conducted on HfO$_2$, there is no comprehensive study on the electronic structure of pure Hf surfaces.

Transition metals and their compounds show wide variety of structures, magnetic, and electronic properties due to their incomplete \textit{d} shells. In 5\textit{d} transition metals and compounds, the 5\textit{d} orbitals are expected to be more extended and the Coulomb repulsion U values are further reduced compared to the 3\textit{d} and 4\textit{d} cases. However, experiments have shown that correlation effects can be important owing to the strong spin-orbit coupling (SOC) of heavy elements \cite{Erickson}.
Furthermore, the interplay between strong SOC and broken inversion symmetry at the surface influences the electronic states and leads to the phenomenon of spin-momentum locking, where the spin of the electrons becomes directly tied to their momentum, resulting in the creation of unique surface states that do not exist in the bulk part of the material. Spin-polarized electrons at the surface are of interest for novel applications in electronics and data processing \cite{lashell1996spin,varykhalov2012ir}. The lack of space-inversion symmetry at the surface can lead as well to topological surface states with helical spin textures induced by strong spin-orbit interaction \cite{hasan2010colloquium}. Elemental metals, for instance Au(111) \cite{yan2015topological},  W(110) \cite{miyamoto2012spin,braun2014exceptional} and Re(0001) surfaces \cite{elmers2020rashba,holtmann2022distinct}  can host Tamm and Shockley states, the two paradigmatic concepts that are used to describe surface states in electronic systems. Further analysis revealed a  \textit{d } orbital character of surface states on W(110) \cite{braun2014exceptional} which is in contrast to the common  \textit{sp} orbital character of surface states (Shockley-type surface states) of topological insulators. 
The question is whether these \textit{d}-like surface states can reveal a significant spin-momentum locking \cite{elmers2020rashba,schemmelmann2021rashba, holtmann2022distinct, schottke2022rashba}.
\newline Here we present both theoretical and experimental characterizations of the electronic structure of the elemental Hf(0001) surface. Our spin-resolved one-step Korringa-Kohn-Rostoker (SPR-KKR) photoemission calculations predict a Rashba split surface state on Hf(0001) similar to that observed on Re(0001) \cite{elmers2020rashba,holtmann2022distinct}. We performed angle-resolved photoemission spectroscopy (ARPES) with both He II and synchrotron radiations to track the electronic properties of the Hf(0001) surface. In particular, the light polarization dependence of ARPES spectra elucidates the orbital character of the electron bands.

\section{Experimental and theoretical methods}

\subsection{Surface preparation}
The hafnium crystal structure is hexagonal with the lattice parameters  \begin{math} a=b= 3.19~ \text{\AA}, ~ c=5.05~\text{\AA} \end{math}, the  reciprocal vector of the (0001) surface is $k_{\mathrm{BZ}}=1.966$~{\AA}$^{-1}$. It belongs to the \begin{math}
    P63/mmc
\end{math} space group (n°194) with the Hf atom lying on the (1/3 2/3 1/4) Wyckoff position. In our experiments, we used a single crystal with the (0001) face produced by the Surface Preparation Laboratory company. 

Due to its high reactivity, the Hf surface cleaning necessitates several cycles of ion bombardment (Ar+) and annealing. The preparation conditions were optimized to an Ar+ ion beam energy of 2 keV during 30 min and the annealing temperature was slowly increased up to 900\textdegree C, maintaining the vacuum conditions in the low 10$^{-9}$ mbar range that yields a high-quality unreconstructed surface, as testified by the (1$\times$1) low-energy electron diffraction (LEED) pattern shown in Figure \ref{Theory} (a). For higher temperatures, above 1000\textdegree C, the surface atoms start to sublimate, the crystal surface is loosing the mirror-like reflectivity and we observed a faint 2$\times$2 reconstruction (See Supplemental Material (SM), Figure S1).

The surface quality was further investigated using X-ray (Mg K$\mathrm{\alpha}$) photoelectron spectroscopy (XPS) identifying  Hf $4f_{7/2-5/2}$ doublet with no trace of the HfO$_{2}$ oxide with a relative binding energy of -3.64 eV \cite{morant1990xps} (See Figure S2). Immediately after the cleaning procedure the intensity of the O 1s peak was negligible, however it was increasing typically after 1 hour of measurements in the analysis chamber with the base pressure in the low 10$^{-10}$ mbar range.
The energy resolution of XPS does not allow to resolve the presence of 4\textit{f} surface peaks, which were clearly detected in He II spectra, as explained more in detail in section III C.

The ARPES experiments were carried out in our home laboratory with the SPECS PHOIBOS 150 hemispherical analyzer and a He~II UV source, and at synchrotron radiation centers SOLARIS (URANOS beamline) and MAX IV (FinEstBeAMS beamline).

\subsection{Computational methods}
  
For theoretical analysis, we used two methods, namely the full-potential linearized augmented plane wave (FLAPW) method as implemented in Wien2K code \cite{Blaha+2001} and the Korringa-Kohn-Rostoker (KKR) Green function method \cite{Ebert+2011} implemented in SPR-KKR code \cite{Ebert+2012}. For the exchange corelation potential we used the GGA functional \cite{PBE+1996}.

In the Wien2K code the Hf(0001) calculation were performed using the slab geometry, each slab consisting of eight layers. The thickness of the slab is 18~\AA\ separated by 14~\AA\ of vacuum. In SPR-KKR the system is projected as a semi-infinite stacking of atomic layers. The KKR Green function calculations were done with the Tight binding mode. (For further details see SM)

The presented theoretical ARPES calculations were derived using the one-step model of photoemission within the SPR-KKR package\cite{braun2018}. The resulting framework, based on the Dirac equation, fundamentally includes all relativistic effects such as Spin-Orbit forces, for instance crucial for reproducing the dichroic phenomena. As well, the geometry and variables of the corresponding experiments are taken into account (e.g. light energy, light polarization, incidence angle) for determining the corresponding matrix-elements. In combination with the tight-binding (TB) model, it was possible to consider both the pure Hf case and the Hf with a layer of oxygen (carbon) on top as contamination. More details on the computational methods can be found in SM.

\section{Results and discussion}

\subsection{Theory vs ARPES spectra } 

\begin{figure*}[ht!]
   \includegraphics[width=1\textwidth]{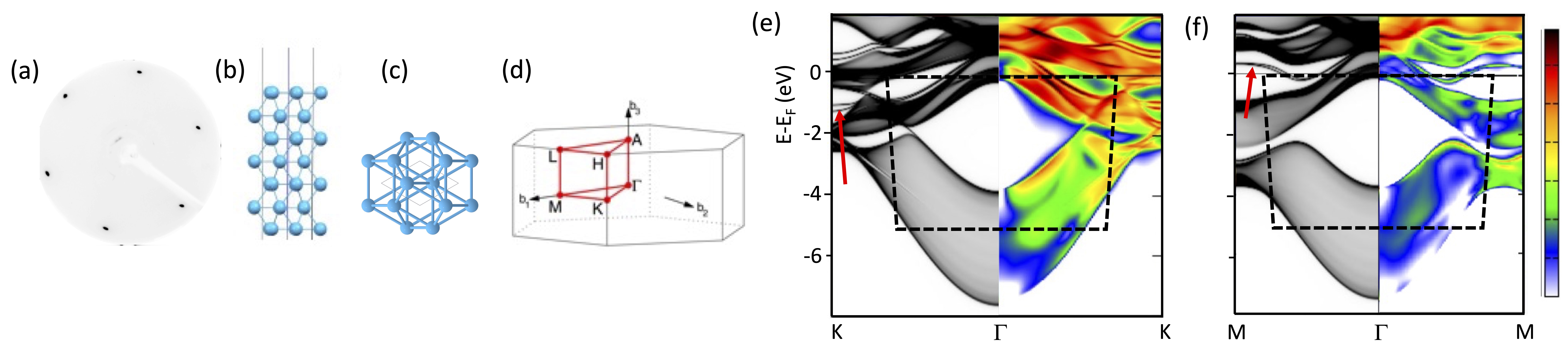}
			\caption{ \label{Theory} (a) LEED pattern obtained with electron beam of 80 eV. (b) Side and (c) top view of pure hafnium crystal structure. (d) The Brillouin zone (BZ) of the hexagonal structure indicating the high symmetry points (e-f)  BSF (left panels) and ARPES (right panels) calculations for the photon energy of 40 eV in the $\overline{\Gamma} \rightarrow \overline{\mathrm{K}}$ and $\overline{\Gamma} \rightarrow \overline{\mathrm{M}}$  high symmetry direction, respectively. Black dotted frames indicate the region measured with He II radiation, shown in Figure \ref{ARPES} (a) and (b).}	
	\end{figure*}

Figure \ref{Theory} presents the side (b) and top (c) view of the Hf(0001) crystal structure and the corresponding Brillouin zone (d). Calculated electron band dispersion by SPR-KKR formalism along the $\Gamma$-K and $\Gamma$-M directions are shown in Figure \ref{Theory} (e) and (f), respectively, displaying the projected states from the $\Gamma$-A high symmetry axis.  The left-hand panels 
represent the Bloch Spectral Function (BSF) and the right-hand ones the ARPES spectra calculated in the one-step model taking into account the experimental geometry. 

The valence band electron configuration of hafnium metal is 5\textit{d}$^{2}$ 6\textit{s}$^{2}$, consequently the calculated dispersion is dominated by hole-like d-electrons crossing the Fermi level and electron-like dispersion of s-electrons, at binding energies between about -2 and -7.5 eV, leaving a large gap, of approximately 3.5 eV at the $\Gamma$ point that progressively closes when approaching K and M high symmetry points.

Remarkable is the presence of a weakly dispersing surface state harbored by the gap in the vicinity of the Fermi level when approaching  the 
$\mathrm{K}$
point (shown by red arrow in Figure \ref{Theory} (e) ). In the $\Gamma$-K direction its dispersion follows the bulk bands and then is crossing the gap in the middle at the K point at the binding energy of about -1 eV, red arrow in Figure \ref{Theory} (e)).
Another surface band can be seen in Figure \ref{Theory} (f), crossing the Fermi level in the ${\Gamma}$ $\rightarrow$ ${\mathrm{M}}$ direction (red arrow). 
As expected, the surface state exhibits strong spin polarization, see Figure S3.

More discussion on the surface states, highlighted by the Bloch spectral function calculations for the upmost Hf atomic layer of crystal in Figure \ref{BSF} (b), will follow in section B.

\begin{figure*}[ht!]
\includegraphics[width=7in]{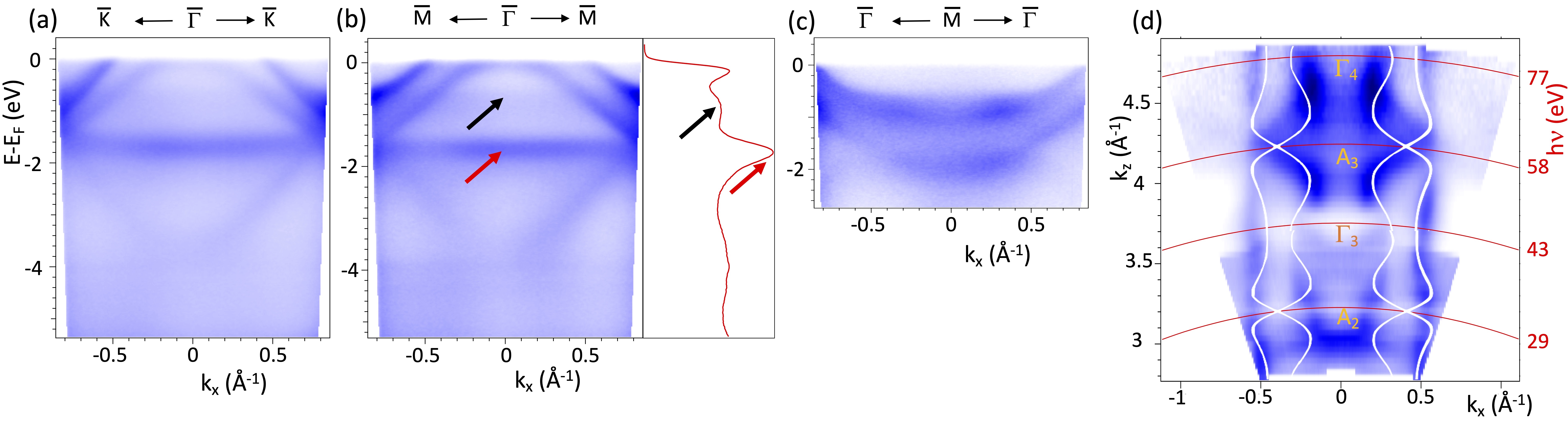}
    \caption{ \label{ARPES} 
    He II ARPES spectra from Hf(0001) surface along the (a) $\overline{\mathrm{K}} \rightarrow \overline{\Gamma} \rightarrow \overline{\mathrm{K}}$ (b) $\overline{\mathrm{M}} \rightarrow \overline{\Gamma} \rightarrow \overline{\mathrm{M}}$ (c) and $\overline{\Gamma} \rightarrow \overline{\mathrm{M}} \rightarrow\overline{\Gamma}$ directions. (d) Photon energy scan from 20 eV to 80 eV shown here at binding energy of -200 meV measured with linearly \textit{p} polarized light. Superimposed white lines represent the ground state calculations. }	
	\end{figure*}

He II ($\mathrm{h\nu}$ = 40.81 eV) ARPES spectra in Figure \ref{ARPES} were measured along the $\overline{\mathrm{K}} \rightarrow \overline{\Gamma} \rightarrow \overline{\mathrm{K}}$  (a), $\overline{\mathrm{M}} \rightarrow \overline{\Gamma} \rightarrow \overline{\mathrm{M}}$  (b) and $\overline{\Gamma}\rightarrow \overline{\mathrm{M}} \rightarrow \overline{\Gamma}$  (c) high symmetry directions. The overall shape of the ARPES spectra fits well the calculations, see Figure \ref{Theory} (e-f) where the black dotted frames indicate the span of He II experimental data. 

Photon-energy dependent ARPES spectra taken with synchrotron radiation light from 20 to 80 eV, [Figure \ref{ARPES} (d)] track the band dispersion along $k_z$ in the $\overline{\Gamma} \rightarrow \overline{\mathrm{A}}$ direction. Here, the photoemission intensity is plotted at E = -200 meV below Fermi level and the white lines drawn on top of the spectrum correspond to bulk ground-state calculations. An inner potential of 14 eV was used to fix the positions in k-space of $\Gamma$ and A points. 
In this set of measurements the cut in the k{$_x$} direction is not oriented in a high-symmetry direction because the sample manipulator does not allow the azimuth sample rotation. The sample misorientation was 17° as seen in the LEED pattern, see Figures \ref{Theory} (a) and S1, i.e. approximately half-way between  $\overline{\Gamma} \rightarrow \overline{\mathrm{K}}$ and $\overline{\Gamma} \rightarrow \overline{\mathrm{M}}$ directions. However, calculations [Figures \ref{Theory} (e) and (f)] and experiment [Figures \ref{ARPES} (a) and (b)] show very similar band dispersion in both, $\overline{\Gamma} \rightarrow \overline{\mathrm{K}}$ and $\overline{\Gamma} \rightarrow \overline{\mathrm{M}}$ directions, so we do not expect signification changes in the k{$_z$} direction either. This is further comforted by a good agreement between experiment and theory shown by white lines in Figure \ref{ARPES} (d). 
Referring to the k{$_z$}-scan we can deduce that the He II photon energy is close to the $\Gamma${$_3$} point of the bulk band structure. This yields the width of the band gap at the $\Gamma$ point, of about 3.8 eV, in agreement with calculations. The gap is limited by 2 hole-like d-bands on the upper side and by s-orbital bands in the lower part and is closing when approaching both M and K high symmetry points according to our calculations.

At the same time, in the experimental data we observe several significant deviations from the theory. 

Firstly, the  surface state predicted by calculations in the $\Gamma$-M direction, supposed to be situated in the gap [see Figure \ref{Theory} (f) and \ref{BSF} (b)] is missing in the experiment. It should appear namely in the M-$\Gamma$-M spectrum, Figure \ref{ARPES} (c).
One explication is that the KKR formalism does not reproduces accurately its energy position and consequently the surface state could be situated in the unoccupied states. 

Secondly, two flat bands, clearly missing in the calculations, are crossing the large gap and are highlighted by arrows in Fig. \ref{ARPES} (b). The energy distribution curve taken at $k{_\parallel}=0$ in the right panel of the figure highlights their presence.
An explanation for their existence is proposed in section B.

  \begin{figure*}[ht!]
   \includegraphics[width=5in]{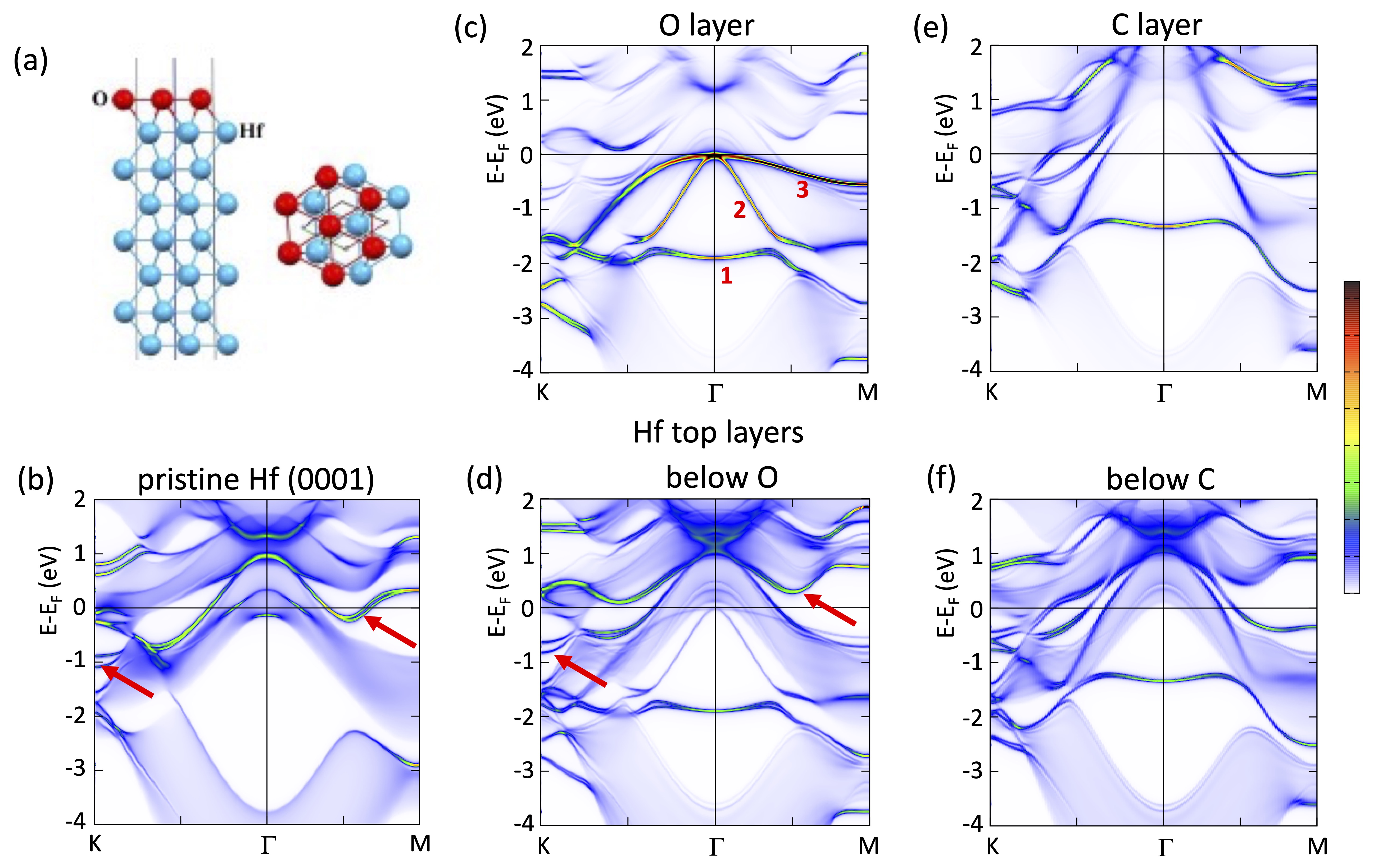}
			\caption{ \label{BSF} (a) Side and top view of the Hf crystal with a monolayer of oxygen (carbon) on the surface. BSF along  ${\mathrm{K}} \rightarrow {\Gamma} \rightarrow {\mathrm{M}}$ for: (b) Hf surface layer on clean Hf(0001), (c) oxygen layer of the Hf+O interface,  (e) carbon layer of the Hf+C interface, and the first hafnium layer below (d) oxygen  and (f) carbon layers. } 	
	\end{figure*}

\subsection{Impact of surface oxygen and carbon on electron states}
The hafnium metal is extremely reactive with oxygen and, as mentioned in section II, the contamination of its surface is unavoidable even in ultra-high vacuum conditions where the most dominant background reactive gas is carbon monoxide. 
In order to simulate the contamination effects we have calculated the surface electron band structure using a simple model in which the Hf(0001) crystal is covered alternatively by a monolayer of oxygen or carbon. Adsorbed atoms are expected to adapt to the crystal structure of the hafnium crystal, as schematically shown in Figure \ref{BSF} (a).

As a reference, Figure \ref{BSF} (b) shows the electron band dispersion of the top atomic layer in the pristine Hf(0001) crystal. Electronic structure of the oxygen and carbon monolayers are presented in Figure \ref{BSF} (c) and (e) respectively, and the corresponding modifications of the upmost Hf layer is shown in Figures \ref{BSF} (d) and [\ref{BSF} (f).

The oxygen ML induces several modifications in the surface electronic bands:

i) The first significant difference appears when comparing the top Hf layer in clean Hf(0001) in Figure \ref{BSF} (b) and the oxygen-covered surface in (d). In the presence of oxygen, the surface state, originally crossing the Fermi level in the $\Gamma$M direction, is shifted to higher energies of about 400 meV (indicated by a red arrow) and is merging with unoccupied bulk bands above the Fermi level.

ii) In the occupied electron states, oxygen atoms generate 3 major bands, labeled 1,2,3 in Fig. \ref{BSF} (c) where we show projected spectral function to the top Hf layers covered by oxygen. Two of them (2 and 3) closely follow the Hf bulk bands. Comparing to projected states on oxygen layer, see Fig. \ref{BSF} (b), we identified a flat band labeled 1 with binding energy -1.8 eV in the BZ center that has significant hybridization traces between Hf 5d electrons and O 2p electrons (Figure S6). The weak band dispersion suggests large localization of the hybridized states near the surface.

The carbon ML  generates a similar flat band in the gap at a lower binding energy (-1.5 eV), see Figures \ref{BSF} (e) and (f)).

The calculations allow for a straightforward interpretation of the photoemission spectra. The oxygen and carbon bands, shown by red and black arrows in the ARPES spectrum, Figure \ref{ARPES} (b), are clearly identified. The presence of these well-defined bands indicates that adsorption of oxygen (carbon) is ordered.  For carbon, the experimental band position is at a lower binding energy (about -0.7 eV), however for oxygen it fits quantitatively, indicating that a 1 ML model is close to the real surface structure.

In this context, it is worth noting that flat bands are unique features of the electronic structure that have recently garnered significant attention as fertile grounds for exploring rare quantum states. In the non-dispersive nature of massive bands, the electrons slow down almost to a halt enabling electronic correlation effects and resulting in ferromagnetism, high-temperature superconductivity or high-temperature fractional quantum Hall effect\cite{Neves2024, Regnault2022}. Flat band-hosting crystal lattices were theoretically proposed almost 40 years ago in the so-called dice structure\cite{Sutherland1986} and later in many other models, such as Kagome, Lieb, pyrochlore or the Penrose tiling. In recent years, this field of physics has been stimulated by the identification of flat electronic bands in 2D moiré heterostructures\cite{Neves2024}. 

Our LEED pattern indicates that the surface of Hf(0001) can, in fact, harbor the dice structure\cite{Sutherland1986, Neves2024, Tassi2024}. The stacking of two top Hf layers and the adsorbed oxygen layer leads to three atoms in the unit cell which is characteristic of the dice lattice\cite{Bae2023}.  So, our observation of the flat band can then be attributed to the adsorption of oxygen (or carbon monoxide) molecules present in the residual gases of the UHV system. The similar system, CO/Cu(111), is used as a model for theoretical studies of electronic properties of the dice structure and it has been shown that CO acts as a repulsive barrier to surface electrons\cite{Tassi2024}. In this case a flat band is appearing at the Fermi level, but in their tight-binding calculation only s-bands have been involved. In our situation p- and d- orbitals are present (see Figure S6) that can modify the biding energy. Moreover, 5d atoms may lead to a flat band with strong spin-orbit coupling\cite{Tang2011}, lifting the spin degeneracy as proven by our calculations in Figure S7 where we observe a clear spin polarization of the flat band of both, the oxygen and the topmost Hf layers.

The shape of the flat bands evolves quickly with time. After several hours they become much wider, supporting the idea that the subsequent layers of oxygen and carbon, or carbon monoxide, are becoming more amorphous in contrast to the well-ordered first layer. It is important to note that at the same time the intensity of the bulk hafnium bands remains stable and is not influenced by an impressive increase of the O 2p peak at 6 eV (see Figure S4). 
As well we draw the reader's attention to the fact that the cross section of the oxygen band in the He II spectra is particularly high, the band's intensity drops significantly in the ARPES spectra of the 2nd Brillouin zone and as well for higher photon energies in synchrotron radiation measurements (Figure S5).
 
A resemblance to a surface state is observable at k{$_x$} = $\pm$ 0.5 {\AA}$^{-1}$ in the k{$_z$}-scan, Figure \ref{ARPES} (d),  where an almost non-dispersing band appears for all k{$_z$} values. However, from the discussion above it seems more plausible that this band is generated by a surface oxygen layer.

\subsection{Hf 4\textit{f} core level shift}

A general approach to core-level binding energies in metals indicates that the surface core-level shift is closely related to the difference in surface energies between the Z and Z+1 elements. In earlier transition metals, the core electrons are more bound at the surface than in the bulk, while for the heavier ones the situation is opposite. This general trend has been interpreted in terms of the bonding-to-antibonding crossing in the filling of the 5d band states when passing the middle of the transition series. \cite{johansson1980core,nilsson1989multielectron}

\begin{figure}[ht!]
   \includegraphics[width=0.4\textwidth]{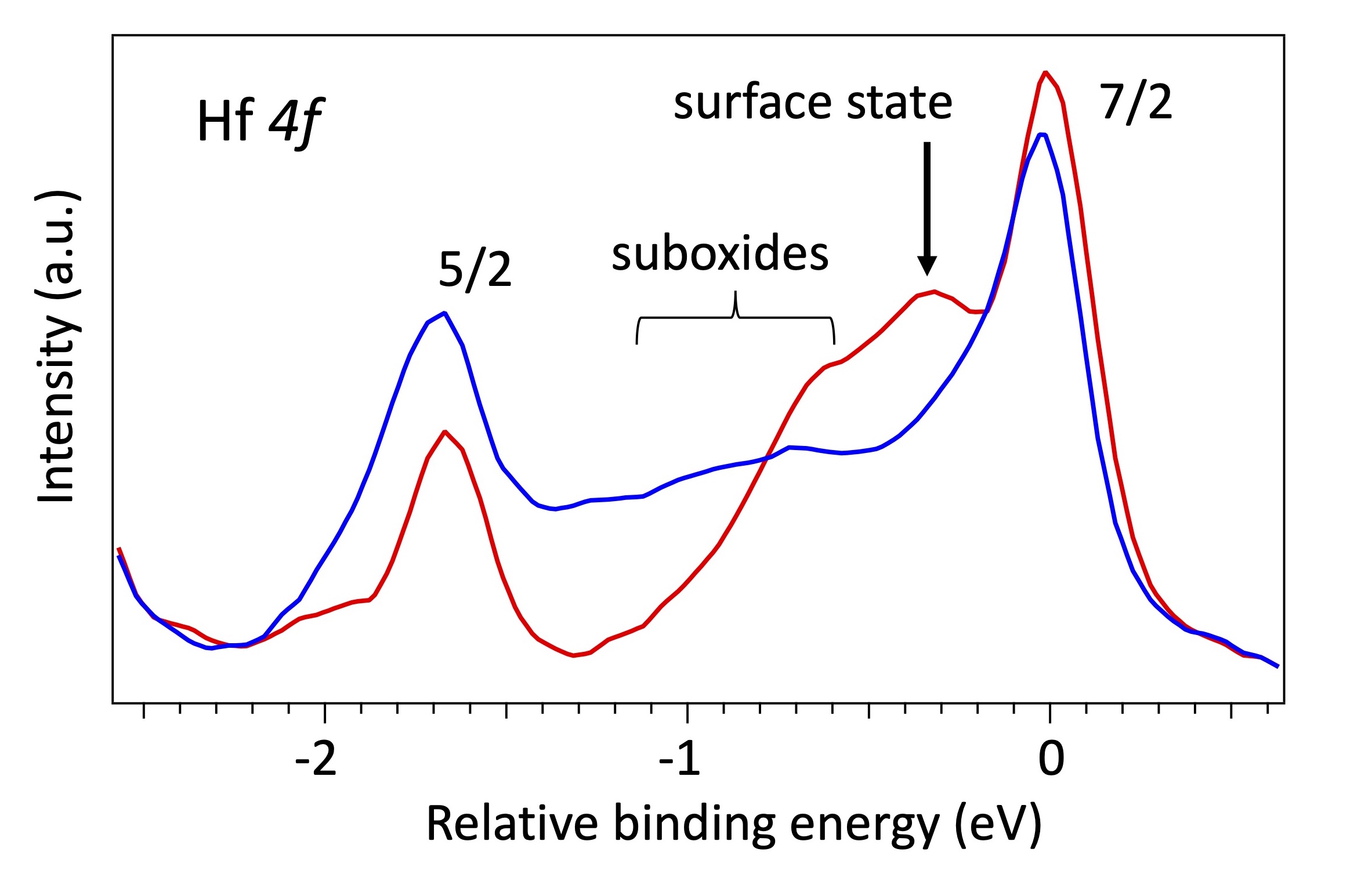}
	
			\caption{ \label{4f} Hf 4\textit{f} spectra measured with He II radiation, red curve indicates spectra taken just after the surface treatment. Blue curve indicates spectra taken after one hour indicating the formation of suboxide.}	
	\end{figure}

\begin{figure*}[ht!]
    \centering
    \includegraphics[width=0.8\textwidth]{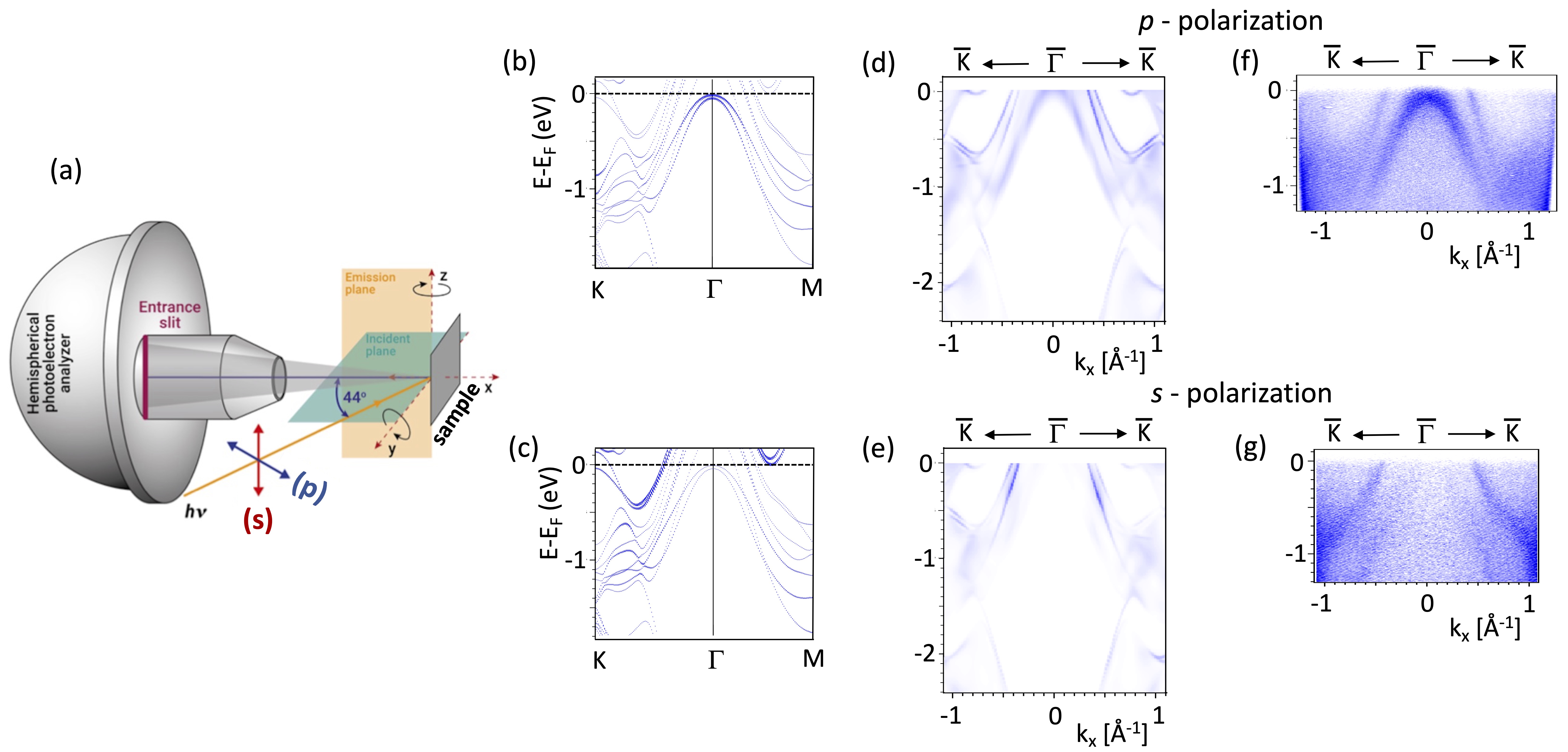}
   \caption{ \label{LD_Hf} (a) Experimental geometry. (b-c) \textit{d}-orbital texture of the Hf(0001) surface band structure with contribution of mainly (b) d$_{z^2}$ and (c) d$_{x^2-y^2}$+d$_{xy}$. (d-e) Corresponding one-step calculation in the ${\Gamma}_4$ point (76 eV) using (d) \textit{p} and and (e) \textit{s} polarizations of the light in the $\overline{\mathrm{K}} \rightarrow \overline{\Gamma} \rightarrow \overline{\mathrm{M}}$ direction respectively. (f-g) ARPES spectra at the same conditions as (d-e).
   } 
\end{figure*}

For Hf 4\textit{f}, the experimental value of the surface core-level shift on the polycrystalline hafnium metal was reported to be -0.42 eV\cite{nyholm1984surface}. However, in the fitting procedure the surface peaks needed to be set significantly broader than the bulk peaks. As concluded by the authors, this was probably caused by both a distribution of surface shifts due to the polycrystalline nature of the samples and a presence of contaminants on the surface.

Our measurements in Figure \ref{4f} allow a more subtle view into the components of Hf 4\textit{f} core levels. On a freshly prepared sample (red spectrum) we can clearly distinguish a surface peak at -0.34 eV from the bulk component, in agreement with theory prediction, - 0.33 eV\cite{johansson1980core,nyholm1984surface}. Even on the best prepared surface in our experimental conditions we observe a shoulder at binding energies of about -0.6 eV and higher, which testifies the presence of a suboxides series. The peak position of the ultimate HfO$_2$ oxide is expected at a binding energy of -3.64 eV \cite{morant1990xps}.
The high reactivity of the Hf clean surface can be observed on the core-level variation. After 1 hour in the UHV conditions, the surface peak disappears almost completely leaving place to a higher intensity of suboxides filling the valley between Hf 4f$_{7/2}$ and Hf 4f$_{5/2}$ peaks, Figure \ref{4f}, blue spectrum.

\subsection{Linear and circular dichroism}

By varying the light polarization in  ARPES experiments, an information on the internal symmetries of a material may be obtained.
Dichroic signals in ARPES depend on both the initial and final states, and can be used 
to uncover the momentum-dependent orbital character in solids \cite{cherepkov1993linear,bansil1999importance}.

\begin{figure*}[ht!]
   \centering
    \includegraphics[width=1\textwidth]{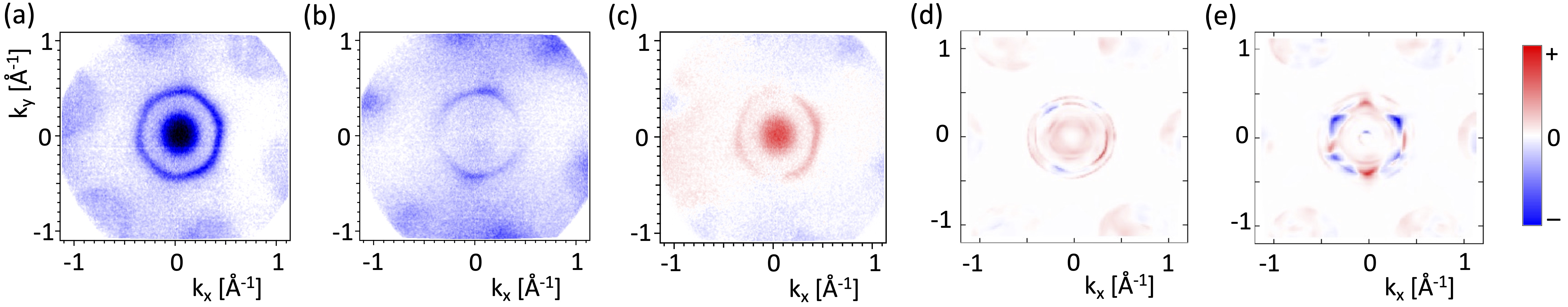}
   \caption{ \label{LD_Hf_exp} (a) Measured constant energy maps at E-E$_\mathrm{F}$= 100 meV using \textit{p} and (b) \textit{s} polarizations of the light. (c) Experimental and (d) calculated linear dichroism (LD=I$_{\textit{s-pola}}$ -I$_{\textit{p-pola}}$). (e) Calculated LD for the Hf(0001) surface covered by one monolayer of oxygen.
   }
\end{figure*}

We discuss here the linear dichroism (LD), the circular dichroism (CD) results are presented in SM, Figure S8. All presented data have been acquired at 70 K. 
The experimental geometry is shown in Figure~\ref{LD_Hf} (a). Figures ~\ref{LD_Hf} (b–c) present theoretical spectra computed using the Wien2k code.The spectral weight of the orbitals is indicated by the thickness of the dotted lines. According to calculations, close to the ${\Gamma}$ point, the inner band mainly exhibits d$_{z^2}$ [Figure~\ref{LD_Hf} (b)] and the outer band d$_{x^2-y^2}$+d$_{xy}$ [Figure~\ref{LD_Hf} (c)] orbital character. 
The d-bands with other symmetries are not shown, as their weight contributions for both light polarizations is much smaller. 
For a better comparison with experimental data one step SPR-KKR calculations have been carried out at the ${\Gamma}_4$ point (photon energy of 76 eV) p- [Figure \ref{LD_Hf} (d)] and s- [Figure \ref{LD_Hf} (e)] light polarizations. Corresponding ARPES spectra in Figures~\ref{LD_Hf} (f) and (g) are in good agreement with the theory and confirm the predicted orbital structure.

According to the selection rules\cite{damascelli2004probing}, since the final state is an even function with respect to the mirror plane (assuming that the photoelectron analyzer is in the mirror plane), orbitals that are odd (even) with respect to the mirror plane can be detected only with light polarization that has the same parity with respect to that plane. This indicates that the orbital character in these two bands corresponds to \textit{d}$_{x^2-y^2}$+\textit{d}$_{xy}$ and \textit{d}$_{z^2}$, which are even with respect to the mirror plane in the $\overline{\Gamma} \rightarrow \overline{\mathrm{M}}$ high symmetry direction, and odd where there is no intensity.

\sloppy The data presented in Figure \ref{LD_Hf_exp} (a, b)  show constant energy maps of Hf(0001) recorded using 76 eV photon energy (${\Gamma}_4$) with \textit{p} (h) and \textit{s} (i) polarizations, respectively. LD calculated as I$_{\textit{s}}$ -I$_{\textit{p}}$ is plotted in Figure \ref{LD_Hf_exp} (c) in a red-blue color scale. The calculated LD is shown in Figure \ref{LD_Hf_exp} (d). Comparing to experimental data we find excellent agreement for both \textit{d}-bands. It is more difficult to look for similarities between calculation and experiment when approaching k-points where experimental factors, such as slight sample misalignment, can play a role. The calculated LD for a model system of Hf(0001) surface covered by one monolayer of oxygen is shown in Figure \ref{LD_Hf_exp} (e). Interestingly, the presence of contamination influences strongly the LD spectrum of the \textit{d}-bands, contrary to what is observed in our measurements. This may suggest that the observed oxygen and carbon bands correspond to a coverage of less than one monolayer.

\section{Conclusion}

The results presented here provide the first experimental ARPES spectra, directly compared to theoretical predictions, offering valuable insights into the electronic structure of the Hf(0001) surface. Theoretical predictions initially indicated the existence of Rashba-split surface states with substantial spin splitting.

ARPES measurements performed with both He II and synchrotron radiation confirmed the theoretical band structures but also revealed the presence of additional flat bands attributed to surface contaminates (oxygen and carbon) at binding energies of -1.8 and -0.5 eV respectively. To corroborate this, we calculated BSF for an oxygen (carbon) monolayer covering Hf(0001). The contaminates induce several modifications of the Hf(0001) surface electronic structure, oxygen atoms generate 3 major bands. Two of them follow closely the Hf bulk band dispersions, the third band is a flat band and is crossing the gap of the $\overline{\Gamma}$ point at -1.8 eV binding energy. Very similar effects are induced by carbon atoms presence namely a flat band is formed in the gap at -0.5 eV. The calculations allowed a straightforward interpretation of the photoemission spectra.

In the light of these findings, the absence of the surface state expected to be reachable in the $\overline{\Gamma} \rightarrow \overline{\mathrm{M}}$ direction 
can now be understood. 
In the presence of oxygen, the surface state, originally crossing the Fermi level in the $\overline{\Gamma} \rightarrow \overline{\mathrm{M}}$ direction, is pushed up in energy. Now, it is fully situated above the Fermi level and its shape is mimicking the hafnium bulk unoccupied bands dispersion.

Remarkable is the observation of the flat band induced presumably by both, adsorption of oxygen and the consequent formation of the dice lattice. In terms of perspective, it would be interesting to study the evolution of this band by controlling the amount of oxygen (carbon monoxide) and by performing spin-resolved photoemission experiments. Manipulating spin polarization in these systems can lead to exotic behaviors, such as, for instance, flat bands with single-spin character\cite{Jugovac2023}.

\section{Acknowledgments}
S.A.K., L.N. and J.M. acknowledge the support by the QM4ST project financed by the Ministry of Education of the Czech Republic grant no. CZ.02.01.01/00/22$_{-}$008/0004572, co-funded by the European Regional Development Fund. 

M.F. acknowledges the support by the project TWISTnSHINE, funded as project No. LL2314 by the ERC CZ Program.

M.C.R., O.H., M.F. and K.H. acknowledge support by the FlatMoi ANR project (No. ANR-21-CE30-0029). 

L.N.-R. was supported by a EUTOPIA PhD Co-tutelle Program.

\clearpage


\title{Supplemental Material for\\ \textit{ARPES studies of Hf(0001) surface: flat bands formation in the dice lattice}}
\maketitle

\section{Theoretical methods}

In the Wien2K code the electron wave function is evaluated as an atomic-like partial wave inside the muffin-tin sphere and as plane wave in the interstitial region. The wave-functions inside the spheres were expanded in spherical harmonics up to the maximum angular momentum $\ell_{\text{max}}^{\text{(APW)}}$. The expansion of the wave functions into plane waves is controlled by the plane wave cutoff in the interstitial region. This cutoff is specified via the product $R_{MT}K_{max}$, where $R_{MT}$ is the smallest muffin-tin (“atomic”) sphere radius and $K_{max}$ is the magnitude of the largest wave vector. In this study we used $\ell_{\text{max}}^{\text{(APW)}}$=10, $R_{MT}K_{max}$=7.0 and $R_{MT}$ were selected as 2.5 a.u. for Hf, 1.8 a.u. for O and C atoms.

The internal geometry of the Hf0001 was optimized using 8~$\bm{k}$-points in the irreducible Brillouin zone (IBZ) distributed according to a $(4\times4\times1)$ Monkhorst-Pack grid\cite{Monkhorst+1976} while the self consistencies of the ground state energies were obtained by 32~$\bm{k}$-points in IBZ, distributed according to a $(8\times8\times1)$ Monkhorst-Pack grid.

In the KKR-Green function calculations, one employs a multipole expansion of the Green function in Voronoi polyhedra so-called Wigner–Seitz sphere for which we used the cutoff $\ell_{\text{max}}^{\text{(KKR)}}$ =3. Take into consideration that the cutoffs $\ell_{\text{max}}^{\text{(APW)}}$ and $\ell_{\text{max}}^{\text{(KKR)}}$ have different meanings within FLAPW and KKR methods, one cannot directly compare their values.

Once the wave functions or the components of the Green function are established, the charge density is derived through the k-space integration over the BZ. In the Wien2k code, the BZ integration was conducted using the modified tetrahedron method \cite{Bloch+1994}. In the SPR-KKR code, the BZ integration was obtained through regular k-mesh sampling, utilizing the symmetry \cite{Huhne+2002}.

\section{Surface preparation}

\begin{figure}[ht!]
    \centering
    \includegraphics[width=0.5\textwidth]{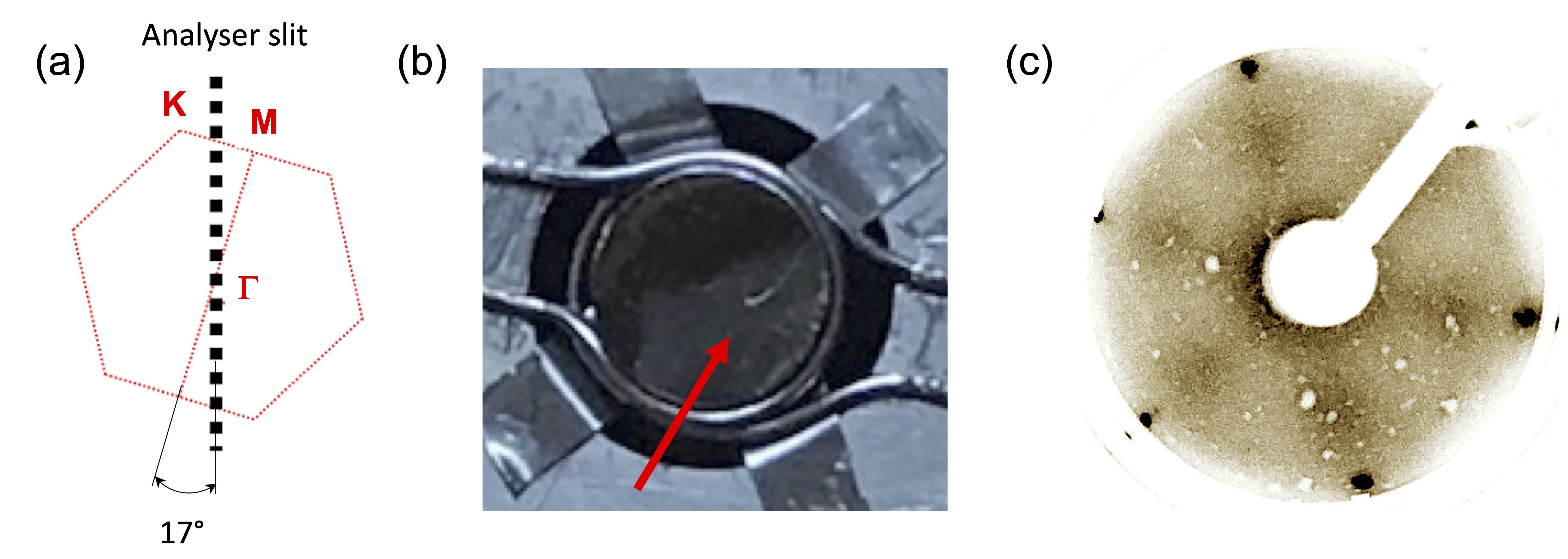}
 \caption{(a) Mechanical mount of the Hf(0001) crystal on the sample plate resulted in a misalignment between the photoemission analyzer slit and the crystal high symmetry axis. (b) A region of the crystal surface (red arrow) damaged after annealing above 1000\textdegree C loosing a mirror-like reflectivity. (c) LEED pattern on the damaged surface showing large 1/2 spots taken with electron beam of 35 eV.}
    \label{fig: SI_Fig1}
\end{figure}

The Hf(0001) was rigidly mounted on a sample plate (Figure \ref{fig: SI_Fig1} (b)). After cleaning the surface, the LEED patterns showed a misalignment between the analyzer slit (vertical in our experimental chamber) and the crystal high-symmetry axis (Figure \ref{fig: SI_Fig1} (a)). The misalignment was corrected by azimuthal rotation of the sample holder, however this could not be done in our series of synchrotron radiation experiments at the URANOS beamline (Solaris), where the mechanical rotation was not available.

Although the data presented were measured on a surface prepared by annealing up to 900°C, we tried a preparation procedure at higher temperatures, above 1000\textdegree C. However, in these conditions, the surface atoms start to sublimate such that a large part of the crystal surface loses its mirror-like reflectivity, as shown by the red arrow in Figure \ref{fig: SI_Fig1} (b). Consequently we observed a faint 2$\times$2 reconstruction, see Figure \ref{fig: SI_Fig1} (c).

\begin{figure}[ht!]
	\centering
	\includegraphics[width=0.5\textwidth]{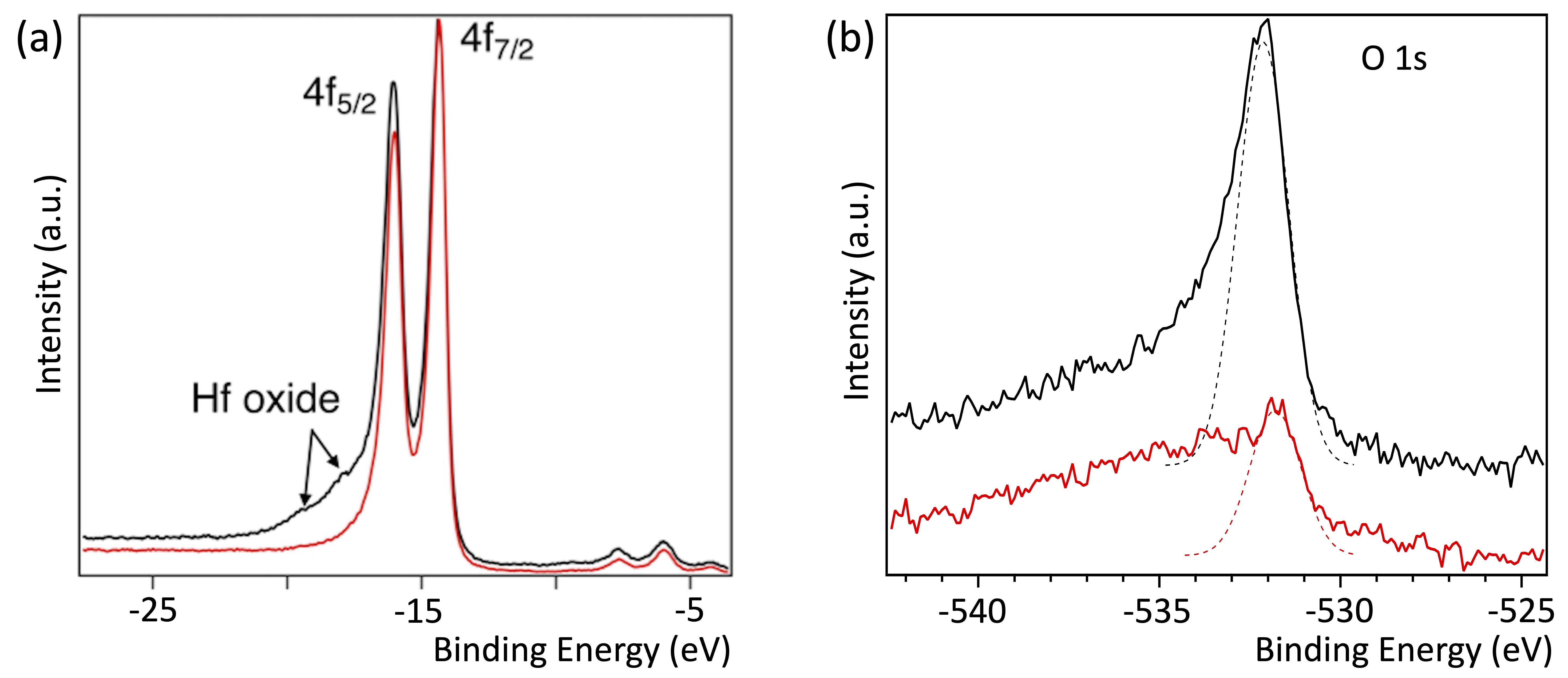}
	\caption{(a) XPS overview spectra before (black) and after (red) one cycle of cleaning. (b) Normalised Hf 4\textit{f} spectra for before (black) and after (red) annealing. (c) Normalised O 1\textit{s} spectra for before (black) and after (red) annealing.}
	\label{fig:Fig2_XPS}
\end{figure}

The surface quality was further investigated using X-ray (Mg K$\alpha$) photoelectron spectroscopy (XPS) identifying  Hf 4\textit{f}$_{7/2, 5/2}$ doublet at -14.36 and -16 eV binding energies. After the the ion bombardment and annealing no trace of the HfO$_{2}$ oxide with a relative binding energy of -3.64 eV \cite{morant1990xps} was detected, red curve in Figure \ref{fig:Fig2_XPS} (a). The energy resolution of XPS does not allow one to resolve the presence of 4\textit{f} surface peaks, which were clearly detected in He II spectra, as explained more in detail in the main text. Increasing contamination is testified by the presence of a new doublet placed at -19.40 and -17.90 eV binding energy corresponds to HfO$_{2}$, black curve in Figure \ref{fig:Fig2_XPS} (a).

Even on freshly prepared surfaces, O 1\textit{s} signal is still present (Figure \ref{fig:Fig2_XPS} (b), red curve) and showing at least two components. The lower binding energy component, schematically indicated by the red dotted Gaussian curve, corresponds presumably to carbon monoxide, the major contaminant of ultra-high-vacuum chambers, not directly bonded to the Hf(0001) surface. The higher binding energy components are attributed to a series of suboxides of hafnium atoms.
The O 1s peak increased considerably after several hours of measurements, as shown by black curve in Figure \ref{fig:Fig2_XPS} (b).

\section{Surface state: spin polarization calculations}

Hafnium is a high-Z element and the surface state is presumably split by the Rashba effect. Our calculations confirm a strong in-plane spin polarization both in the $\overline{\mathrm{\Gamma}} \rightarrow \overline{\mathrm{M}}$ and $\overline{\mathrm{\Gamma}} \rightarrow \overline{\mathrm{K}}$ directions, as shown in Figure \ref{fig: SI_Fig3_spin} (b). We note that in the $\overline{\mathrm{\Gamma}} \rightarrow \overline{\mathrm{K}}$ case, a strong out-of-plane spin polarization is present.

\begin{figure}[ht!]
    \centering
    \includegraphics[width=0.4\textwidth]{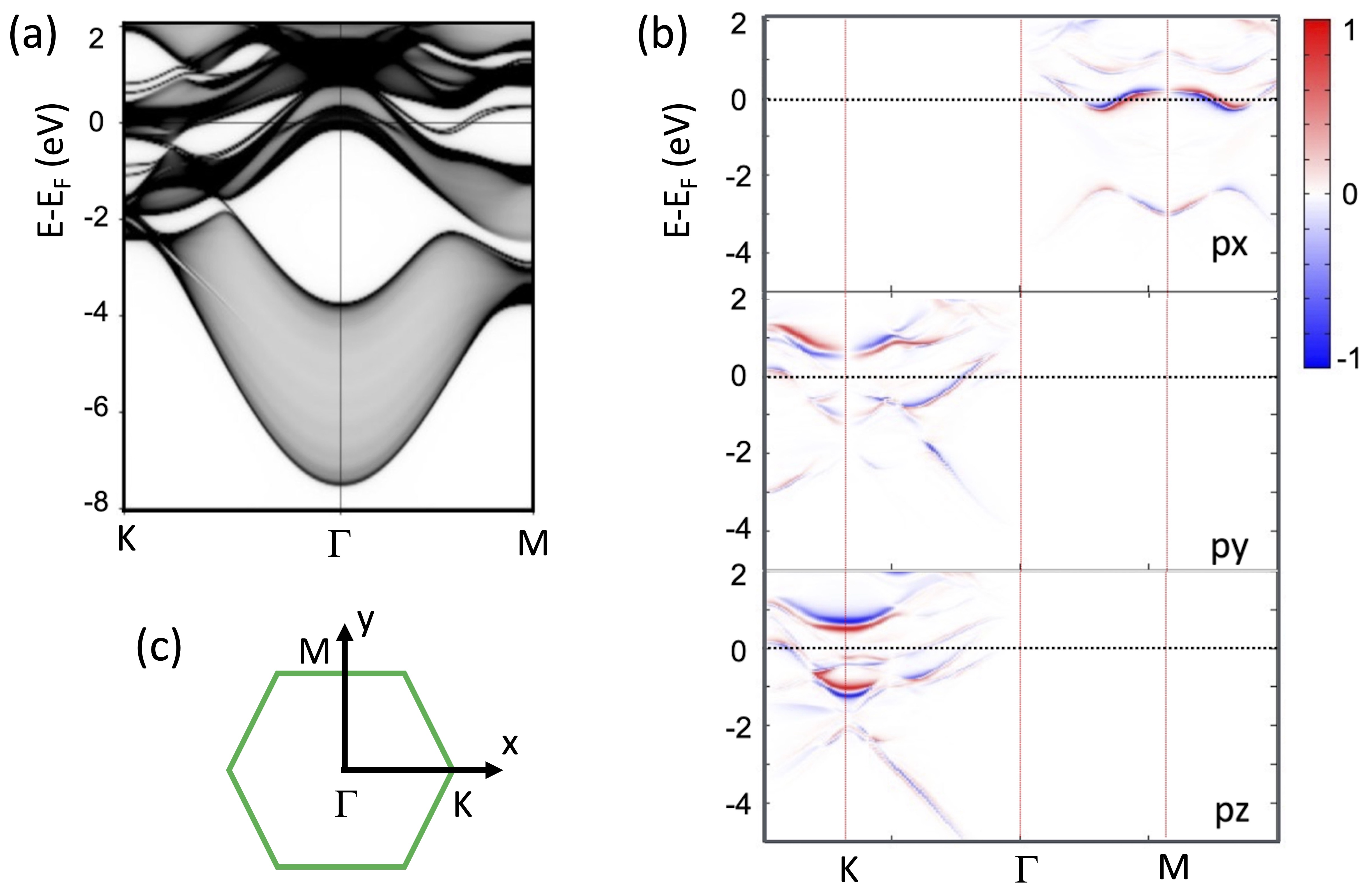}
    \caption{(a) BSF along  ${\mathrm{K}} \rightarrow {\mathrm{\Gamma}} \rightarrow {\mathrm{M}}$ direction. (b) \textit{p}$_x$, \textit{p}$_y$ and \textit{p}$_z$ components of the spin polarization. (c) A scheme of the coordinates used in the calculations. }
    \label{fig: SI_Fig3_spin}
\end{figure}

\section{Impact of surface oxygen and of carbon on electron states}

 Despite rapid contamination of the surface under vacuum, the intensity of the bulk Hf bands remains relatively stable and is not influenced by the increase in the peak of O 2p at 6 eV as seen in the intensity spectra in Figure \ref{fig:Overview}.  

\begin{figure}[ht!]
    \centering
    \includegraphics[width=0.9\linewidth]{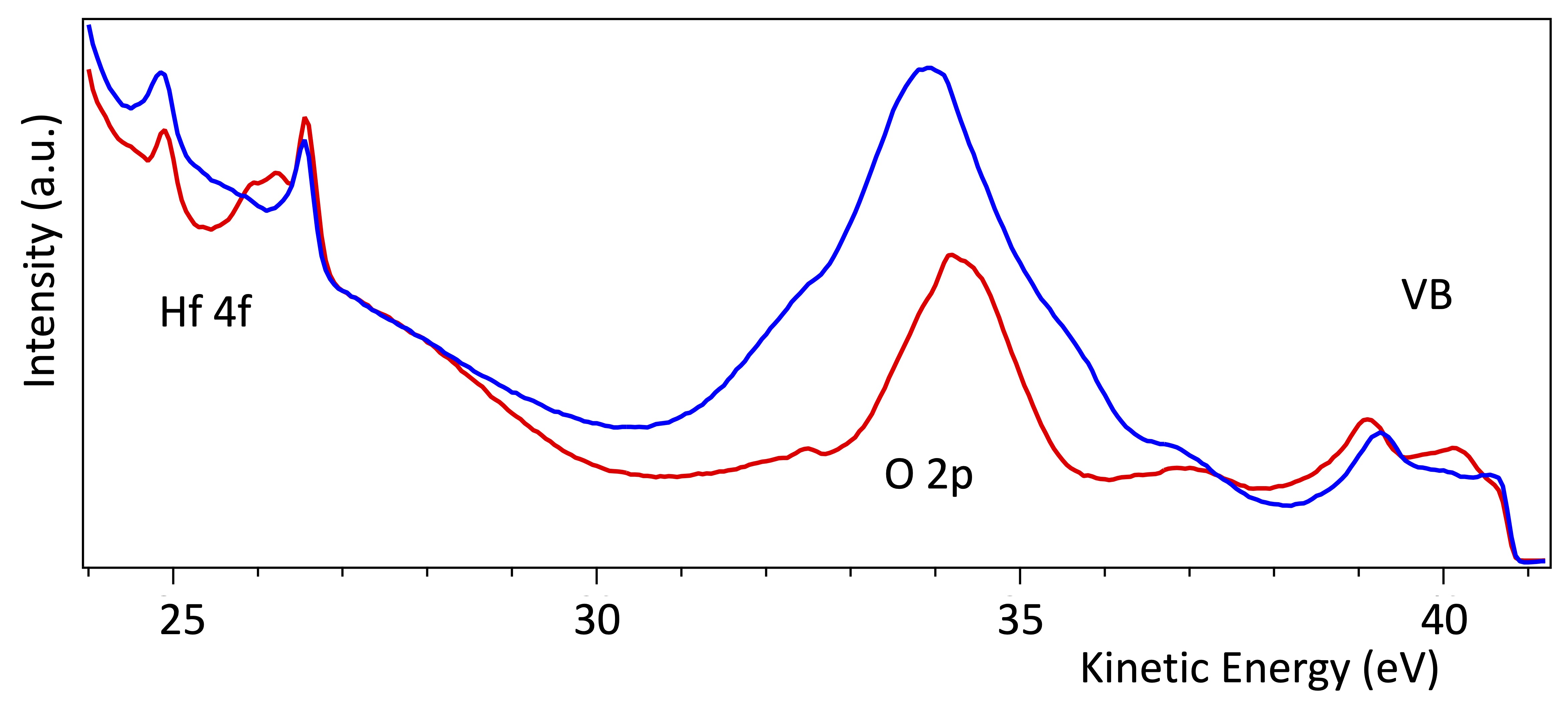}
    \caption{Intensity profiles of the He II spectra in the $\overline{\mathrm{\Gamma}}$-$\overline{\mathrm{M}}$ direction after the cleaning (red spectrum) and after approximately 2 hours in the vacuum (blue spectrum).}
    \label{fig:Overview}
\end{figure}

The shape of the flat band evolves with time. Typically after several hours, it becomes much wider, as evidenced from the comparison between Figures \ref{fig: SI_Fig4_O} (a) and (b). This supports the idea that thicker layers of CO molecules are becoming more disordered in contrast to the well-ordered first layer.
It is also important to note that the cross section of the oxygen band plays a role.  
The cross section is particularly high for lower photon energies, especially for He II, and drops when going towards soft X-ray regime, see Figure \ref{fig: SI_Fig4_O} (b), where spectra measured with synchrotron radiation at two different photon energies  are shown. At the same time the intensity of the hafnium bulk bands remains approximately stable and does not show a significant modulation due to the oxygen and carbon presence.

\begin{figure}[!ht]
    \centering
    \includegraphics[width=0.5\textwidth]{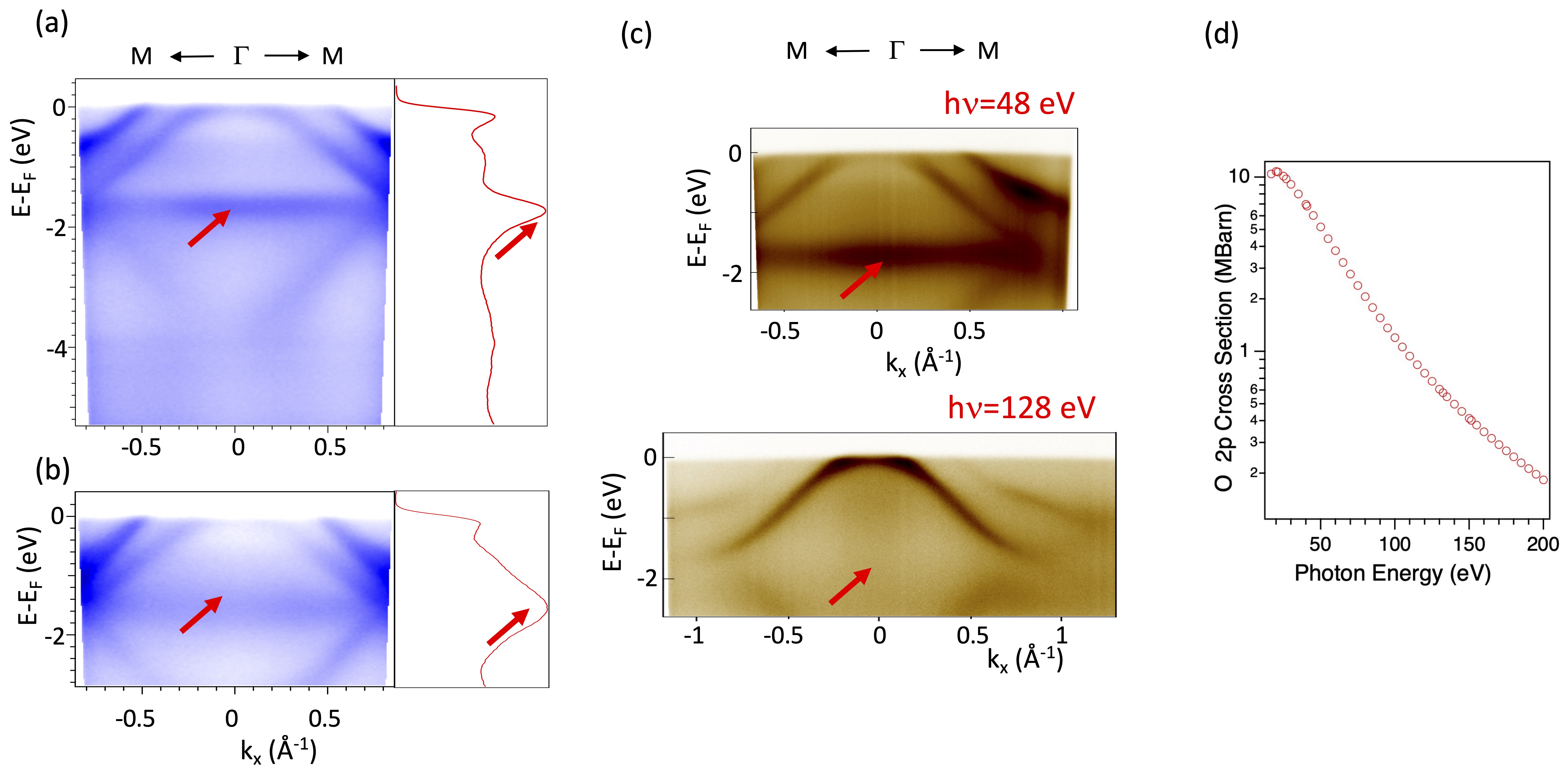}
    \caption{(a) He II spectrum (${\mathrm{\Gamma}}$$_3$ point) in the $\overline{\mathrm{\Gamma}}$-$\overline{\mathrm{M}}$ direction showing the oxygen (red arrow) flat band. (b) The same spectrum after approximately 2 hours in the vacuum. The oxygen flat band is much wider as seen in the intensity profile, right panel. (c) Spectra at two different photon energies taken with synchrotron light. (d) Calculated photon energy dependence of the photoionization cross-sections for O 2\textit{p} atomic levels, reproduced from ref \cite{yeh1985atomic}.}
    \label{fig: SI_Fig4_O}
\end{figure}

\section{Oxygen flat band}

\begin{figure}[ht!]
    \centering
    \includegraphics[width=0.5\textwidth]{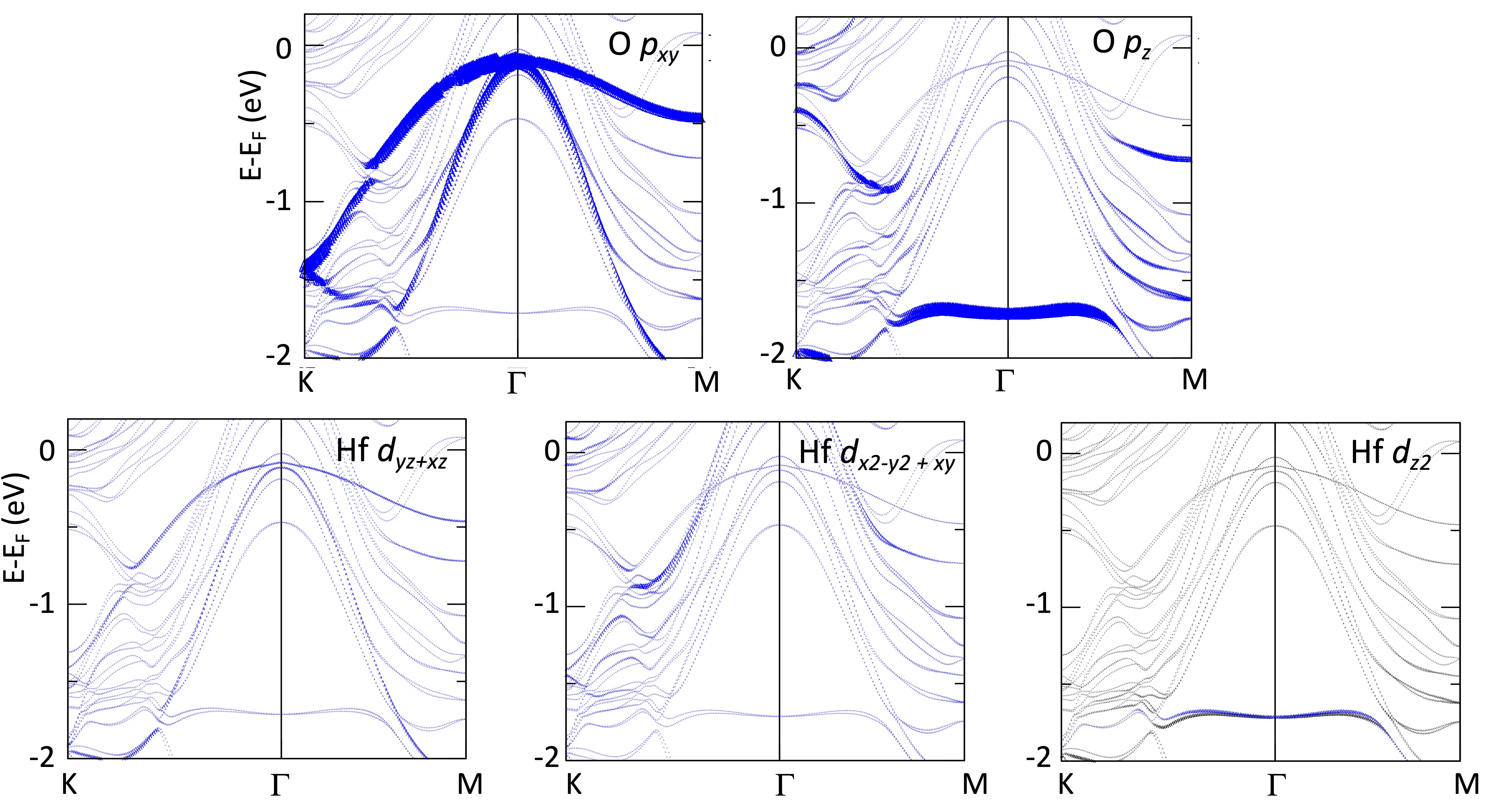}
   \caption{Orbital texture of the oxygen layer (upper row) and of the topmost hafnium layer below the oxygen (bottom row). }
   \label{fig: O_orbitals}
\end{figure}

The calculations clearly indicate that O \textit{p} orbitals are involved in the flat band (Figure \ref{fig: O_orbitals}) that are preponderantly oriented in the \textit{z} direction . The major contribution comes from hybridization between O \textit{p}$_{z}$ and Hf \textit{d}$_{z^2}$ orbitals. 

In contrast, and as expected, the spin polarization of the flat band is contained in the plane of the sample surface and is generated by the \textit{p}$_{x}$ and \textit{p}$_{y}$ orbitals of both, O and top most Hf atomic layers, see Figure \ref{fig: FB_spin}.

\begin{figure}[ht!]
    \centering
    \includegraphics[width=0.5\textwidth]{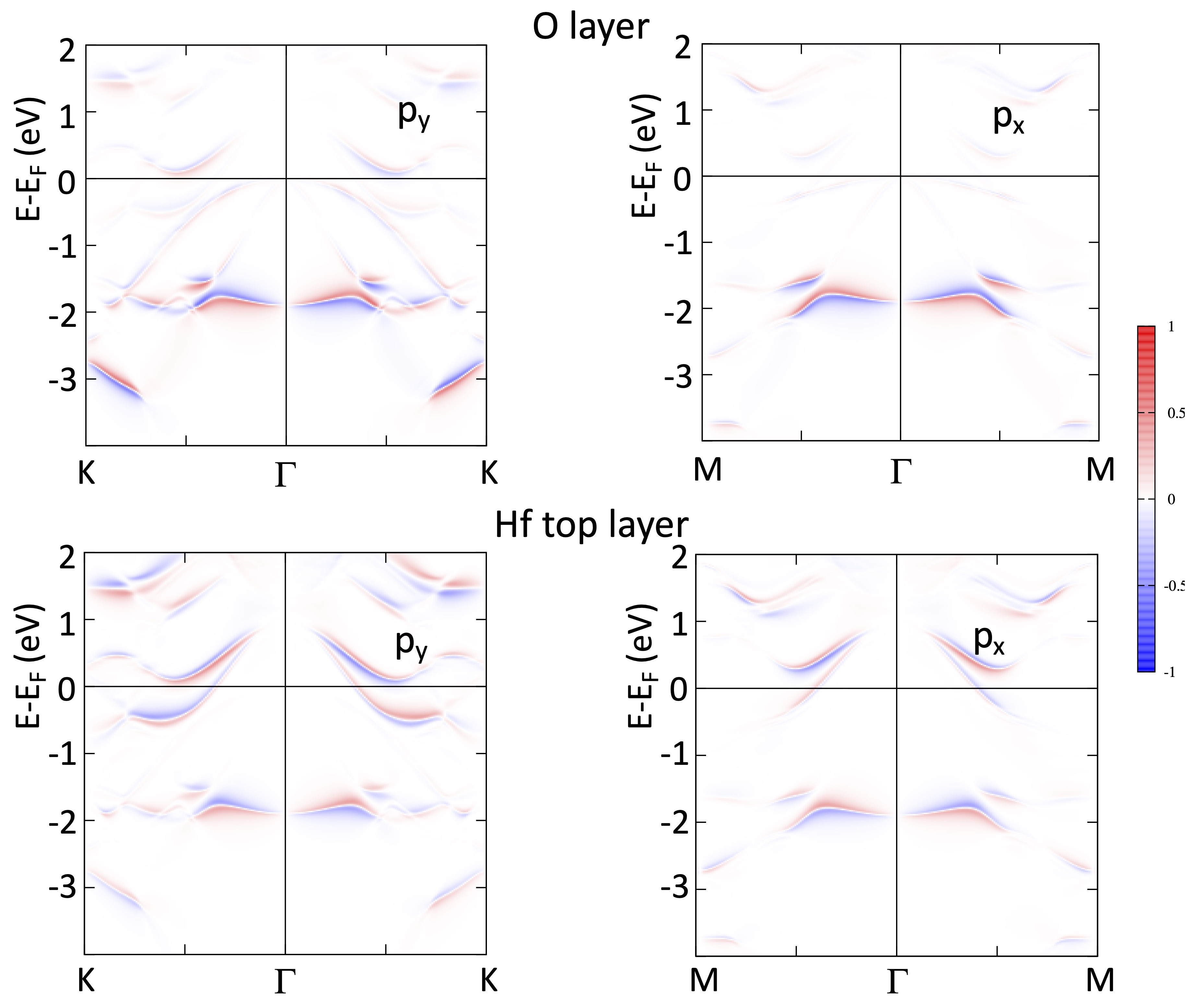}
   \caption{Calculated spin polarization of the oxygen layer (upper row) and of the topmost hafnium layer below the oxygen (bottom row). A scheme of the coordinates used in the calculations is shown in Figure S\ref{fig: SI_Fig3_spin} (c). }
     \label{fig: FB_spin} 
\end{figure}

\section{Circular dichroism}

We have also performed the circular dichroism (CD) ARPES Fermi surface mapping using circular right and circular left (CR/CL) light polarizations at a photon energy of 76 eV, corresponding to $\mathrm{\Gamma_{4}}$ level. The constant energy cut at E-E$_\mathrm{F}$=-100 meV using CR is shown in Figure \ref{CD} (a), while the extracted CD image is shown in Figure \ref{CD} (b). The magnitude of the circular dichroism is given by CD = I$_{\mathrm{CR}}$- I$_{\mathrm{CL}}$ where I$_{\mathrm{CR}}$ and I$_{\mathrm{CL}}$ are the photoemission intensities obtained for right and left circularly polarized lights, respectively. The CD plot highlights regions with significant dichroism, which arises from the interaction of circularly polarized light with the electronic states of the Hf(0001) surface. The differences in intensity patterns between CR and CL polarization provide insights into the orbital character of the electronic states. 

The calculated CD for a model system of Hf(0001) surface covered by one monolayer of oxygen is shown in Figure \ref{CD} (d). Again, as it was the case for LD, the presence of contamination influences strongly the CD spectrum of the d-bands that is not observed in 
measurements.

\begin{figure}[ht!]
    \centering
    \includegraphics[width=0.5\textwidth]{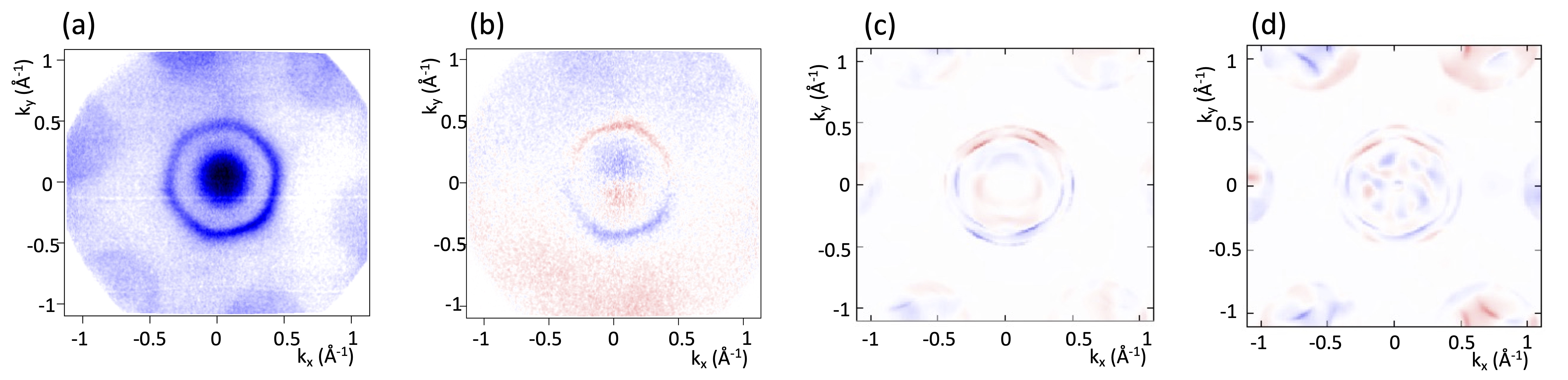}
   \caption{ \label{CD} (a) constant energy map at E-E$_F$= 100 meV using RCP polarization, (b) experimental CD (c) calculated CD, (d) calculated CD for Hf(0001) covered by one monolayer of oxygen. }
\end{figure}

\clearpage
\bibliographystyle{plain}
\bibliographystyle{unsrt}


\end{document}